\documentclass [prb,aps,twocolumn,letterpaper,superscriptaddress] {revtex4}

\usepackage {graphicx}
\DeclareGraphicsExtensions {.eps}

\begin{document}

\title {Synthesis and characterization of atomically-thin graphite films on a silicon carbide substrate}

\author {E. Rollings}
\affiliation {Department of Physics, University of California, Berkeley, CA 94720, USA}

\author {G.-H. Gweon}
\affiliation {Department of Physics, University of California, Berkeley, CA 94720, USA}

\author {S. Y. Zhou}
\affiliation {Department of Physics, University of California, Berkeley, CA 94720, USA}

\author {B. S. Mun}
\author {J. L. McChesney}
\affiliation {Advanced Light Source, Lawrence Berkeley National Laboratory, Berkeley, CA 94720, USA}

\author {B. S. Hussain}
\affiliation {Materials Sciences Division, Lawrence Berkeley National Laboratory, Berkeley, CA 94720 USA}

\author {A. V. Fedorov}
\affiliation {Department of Physics, University of California, Berkeley, CA 94720, USA}
\affiliation {Advanced Light Source, Lawrence Berkeley National Laboratory, Berkeley, CA 94720, USA}

\author {P. N. First}
\affiliation {School of Physics, Georgia Institute of Technology, Atlanta, GA 30332-0430}

\author {W. A. de Heer}
\affiliation {School of Physics, Georgia Institute of Technology, Atlanta, GA 30332-0430}

\author {A. Lanzara}
\affiliation {Department of Physics, University of California, Berkeley, CA 94720, USA}
\affiliation {Materials Sciences Division, Lawrence Berkeley National Laboratory, Berkeley, CA 94720 USA}

\date {\today}

\begin {abstract}

This paper reports the synthesis and detailed characterization of graphite thin films produced by thermal decomposition of the (0001) face of a 6H-SiC wafer, demonstrating the successful growth of single crystalline films down to approximately one graphene layer. The growth and characterization were carried out in ultrahigh vacuum (UHV) conditions. The growth process and sample quality were monitored by low-energy electron diffraction, and the thickness of the sample was determined by core level x-ray photoelectron spectroscopy. High-resolution angle-resolved photoemission spectroscopy shows constant energy map patterns, which are very sharp and fully momentum-resolved, but nonetheless not resolution limited. We discuss the implications of this observation in connection with scanning electron microscopy data, as well as with previous studies.

\end {abstract}


\maketitle

Novel carbon-based structures represent key materials for new technological applications and devices.  The wave of activity in the past few years surrounding carbon nanotubes 
[see, e.g., Ref.'s \onlinecite {javey, tans, bachtold, snow}], which can be thought of as graphene sheets cut and rolled up into nanometer-sized cylinders, has brought renewed interest in graphite and related materials. In particular, attention has turned most recently to planar graphene sheets themselves as low-dimensional graphitic structures whose remarkable electronic transport properties present great scientific and technological potential \cite {berger, novoselov}.

Another factor that makes graphene layers interesting is the fact that, despite many decades of graphite research, high quality single crystalline graphite samples are very difficult to obtain. In fact, the most common form of graphite appearing in the literature is highly-oriented pyrolytic (HOPG) graphite [see, e.g., Ref.'s \onlinecite {zhou, pivetta, hoffman, skytt, novoselov}], which is partially polycrystalline due to its complete in-plane azimuthal disorder. This may be attributed to the fact that natural graphite single crystals are both rare and known to be inhomogeneous \cite {charrier}. Similarly, kish graphite is even more defect-prone than HOPG \cite {kopelevich}. These facts make it appealing to study synthetic single crystalline graphite films produced in UHV, grown either by chemical vapor deposition \cite {oshima} or by thermal desorption of Si from a 6H-SiC substrate \cite {bommel, forbeaux, charrier, berger}.

In this paper, we report a detailed characterization of samples grown by the latter method. A unique aspect of this study is that all characterization tools were UHV techniques. Most notably, we present the first quantitative characterization of the number of graphene layers using core level x-ray photoelectron spectroscopy (XPS), as well as the first constant-energy electron density maps near Fermi energy ($E_F$) using high-resolution angle-resolved photoemission spectroscopy (ARPES)\@.  These measurements, in addition to low-energy electron diffraction (LEED) and scanning electron microscopy (SEM) measurements, provide crucial information on both the thickness of the graphene overlayer and the uniformity of the crystal surface. To our knowledge, there is only one previous study quantitatively characterizing the number of graphene layers on samples grown by this method \cite {charrier}, and hard X-ray diffraction was the technique used in that study. Our results, in general in good agreement with previous work \cite {forbeaux, charrier, berger}, demonstrate that it is now possible to grow, characterize, and study ultrathin single crystal graphite samples, ranging from 1-2 to 3-4 graphene layers, in a single UHV environment.

Graphene films were produced on the Si-terminated (0001) face of an n-type 6H-SiC single crystalline wafer purchased from Cree, Inc. Sections roughly 4 mm $\times$ 8 mm in size were cut from the wafer using a diamond saw, and cleaned with acetone and isopropyl alcohol in an ultrasonic bath. These samples were then mounted with Ta-foil clips to a Mo sample holder and entered into an ultrahigh-vacuum preparation chamber (base pressure 1e-10 Torr). Before growing graphene layers, the sample was cleaned {\em in situ} by annealing up to 850 $^\circ$C under silicon flux for 20 to 30 minutes. The silicon flux was produced by a silicon ingot heated by electron bombardment to approximately 1200 $^\circ$C and was calibrated with a quartz crystal monitor (deposition rate about 3 \AA/min). This procedure removes native surface oxides by the formation of volatile SiO, which sublime at this temperature \cite {bermudez}.
 
The growth process was monitored by LEED patterns, which were obtained at room temperature. XPS measurements were carried out using Mg K$\alpha$ radiation and normal emission geometry. ARPES measurements were conducted at beam line 12 of the Advanced Light Source (ALS) at Lawrence Berkeley National Lab. An SES 100 electron analyzer was used in each case. The base pressure of the XPS (ARPES) chamber was 1e-9 (3e-11) Torr. The measurement temperature was 300 K for XPS and 15 K for ARPES\@. For the XPS measurements, no sample preparation was necessary. For the ARPES measurements, the sample was cleaned by annealing at temperatures steadily increasing to $\approx 700 ^\circ$C in high vacuum (better than 8e-8 Torr) for several hours.  The data reported below are mainly from three samples, henceforth referred to as samples A, B, and C.  These samples are labeled in order of increasing thickness, a parameter controlled primarily by variation in the annealing temperature \cite {berger}.

\begin{figure}[t]

\includegraphics*[width=2.5 in]{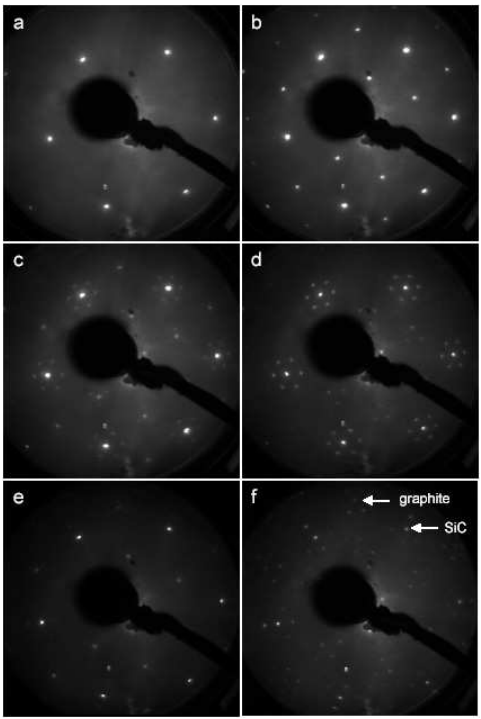}

\caption {(a-d) LEED patterns with a primary energy of 180eV, obtained at four different stages during the growth of sample A.  (a) 1$\times$1 spots of SiC, after a 5 min anneal around 1000 $^\circ$C followed by the initial cleaning procedure under Si flux. (b) ($\sqrt3\times\sqrt3$)R30 reconstruction, after 5 min around 1100 $^\circ$C. (c) ($6\sqrt3\times6\sqrt3$)R30 reconstruction, after 10 min around 1200 $^\circ$C. (d) Sharper($6\sqrt3\times6\sqrt3$)R30 pattern, after 4 min around 1250 $^\circ$C. (e) and (f) LEED patterns with a primary energy of 130eV taken at the same stages as (c) and (d), respectively.  In (f), the appearance of the 1$\times$1 spots of graphite, located slightly farther out from the center relative to the spots observed at similar positions in (b, c, and e), indicates that a thin graphene overlayer has been formed in this last step.}

\label {fig1}

\end{figure}

Figure 1 shows selected LEED patterns obtained at different stages during the growth of sample A\@. After the initial cleaning procedure under Si flux, the LEED pattern (not shown) displayed a 3$\times$3 reconstruction with respect to the SiC substrate, as has been well documented in the literature \cite {bermudez, kaplan, starke}. A subsequent 5 minute annealing at $\approx 1000 ^\circ$C in the absence of Si flux gives rise to the sharp pattern shown in panel (a), corresponding to the 1$\times$1 spots of SiC\@. Further annealing for 5 min at $\approx 1100 ^\circ$C produces the ($\sqrt3\times\sqrt3$)R30 reconstruction shown in panel (b), attributed to a structural model comprised of 1/3 layer of Si adatoms in threefold symmetric sites on top of the outermost SiC bilayer \cite {owman, northrup}. Finally, annealing for 10 min at $\approx 1250 ^\circ$C produces the complex ($6\sqrt3\times6\sqrt3$)R30 pattern shown in panel (c). Notably, a slight increase of the annealing temperature gives rise to the more well developed ($6\sqrt3\times6\sqrt3$)R30 pattern shown in panel (d). Panels (e) and (f) show LEED patterns obtained at a lower electron energy, under the same conditions as panels (c) and (d), respectively.  Graphite diffraction spots clearly observed only in panel (f) indicate that a thin graphene overlayer is formed in the last step.

\begin{figure}[b]

\includegraphics*[width=2.5 in]{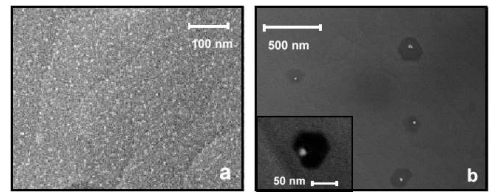}

\caption {SEM images recorded on the surface of a sample grown in the same conditions as sample A. (a) Typical region of the surface, showing a pattern with a length scale on the order of hundreds of nanometers, and a finer pattern with a length scale on the order of 10 nm. (b) Region marked by a cluster of unidentified structures characterized by six-sided geometry, reflecting the underlying hexagonal lattice.}

\label {fig2}

\end{figure}

Figure 2 presents SEM images of a sample very similar to sample A\@. The image shown in panel (a) exhibits two patterns, one in a large length scale, corresponding to hundreds of nm, and another in a small length scale, on the order of 10 nm.  The small length scale coincides with a typical terrace size observed by scanning tunneling spectroscopy \cite {charrier}, while the large length scale may correspond to the single-crystal grain size.  The image shown in panel (b) exhibits a similar pattern and, in addition, unidentified structures - white specks each enclosed by a large black region. Spread out widely over the sample surface, the white specks and enclosing black regions show facets reflecting the underlying hexagonal lattice. An investigation into potential causes for the formation of such structures is currently under way.

Figure 3 shows XPS spectra of the C 1s and Si 2p core levels and their line shape fits.  For all samples, the Si 2p spectrum consists of a single peak located at $\approx$ 101.6 eV, attributed to the SiC bulk. The C 1s spectrum consists of three components, labeled as X, G, and S\@.
In the analysis that follows we will make use of the latter two, which are assigned to the graphite overlayer (G) and the SiC bulk (S)\@.  These assignments are in agreement with previous work \cite {simon, johansson}. In particular, the small chemical shift of peak S relative to pure SiC by about 0.4 eV has been noted before, and was attributed to a Fermi level pinning associated with surface metallization \cite {johansson, simon, soe}. Peak X, broad and weak, is of a less clear origin. Previously, a similar feature was interpreted as arising from C-C bonding in a Si-depleted interfacial region between graphite and SiC \cite {simon, johansson}.
 
\begin{figure}[t]

\includegraphics*[width=3.0 in]{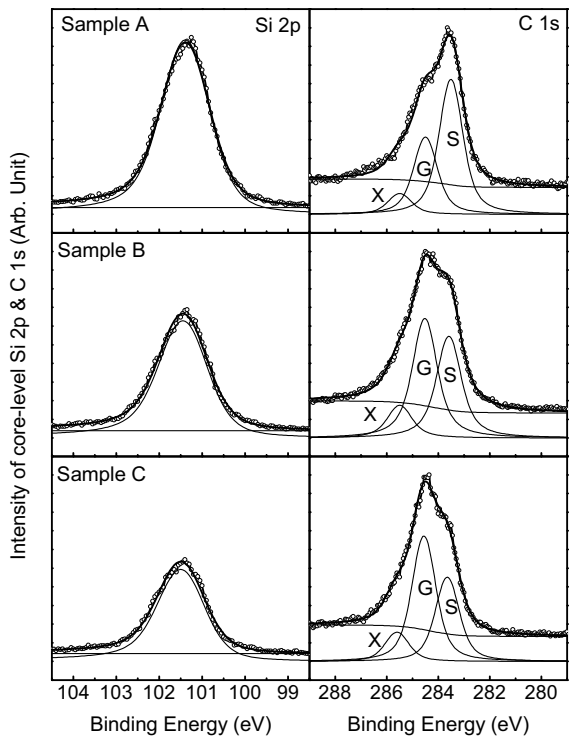}

\caption {XPS results showing data (connected symbols) and their fits (solid lines) for samples A, B, and C.  Individual Voigt functions and Shirley background are shown as lines as well.  In the C 1s spectra, peaks G and S are identified with the graphite overlayer and SiC bulk.  In each row, the two panels are plotted using a common intensity scale.}

\label {fig3}

\end{figure}

Using the relative intensities of three peaks (Si 2p, G, and S), we obtained information on the thickness of the graphite overlayer for each of the three samples studied. Assuming the simple model of a semi-infinite substrate with uniform overlayer of thickness $t$, one finds that the ratio of the intensity $N_G$ of the graphite peak to the intensity $N_R$ of a reference peak (either Si 2p or S) is related to $t$ by the equation \cite {fadley, manella},
\[
\frac {N_G} {N_R} = \frac {T (E_G) \rho^\prime C_G \Lambda^\prime (E_G) [1 - \exp (-t /\Lambda^\prime (E_G))]} {T (E_R) \rho C_R \Lambda (E_R) \exp (-t / \Lambda^\prime (E_R))} \times F
\]
where $E$ is the kinetic energy of photoelectrons associated with a given peak, $T$ is the transmission function of the analyzer, $C$ is the differential crossection ($d\sigma / d\Omega$), $\rho$ is the atomic density, $\Lambda$ is the inelastic mean free path, $F$ is a geometrical correction factor due to photoelectron diffraction, and the superscript $^\prime$ indicates quantities referred to the graphene overlayer as opposed to the SiC bulk\@. Mean free paths were estimated using the so-called TPP-2M formula \cite {tanuma, tougaard}.  Differential cross sections were calculated using tabulated values \cite {yeh} for total cross sections and asymmetry parameters.  The factor $F$ was computed using x-ray photoelectron diffraction data in the literature collected on similar samples \cite {simon}.

\begin{table}[t]
\begin{tabular}{|c||c|c|c|}
\hline Reference & Sample A & Sample B & Sample C \\
\hline \hline Si 2p & 1.2 $\pm$ 0.6 & 2.4 $\pm$ 1.2 & 2.9 $\pm$ 1.5 \\
\hline C 1s (S) & 1.5 $\pm$ 0.5 & 2.8 $\pm$ 0.8 & 3.4 $\pm$ 1.0 \\
\hline
\end{tabular}
\caption {Sample thickness (ML) deduced from XPS analysis, using two different reference peaks from the substrate.}
\end{table}

Solving the above equation for thickness $t$ and dividing by an interlayer
spacing of 3.35 \AA 
gives the results summarized in Table 1. The excellent agreement between values obtained using two different reference peaks shows that this method works well.  Note that using the C 1s peak (S) as a reference is obviously the preferred method, since in this case several factors ($C$, $T$, and $F$) simply drop out of the calculation to within a good approximation.
Indeed, the systematic deviation up to $\approx 20$ \% when the Si 2p peak is used as a reference can be attributed to the uncertainty of these factors, most probably that of $C$\@. In any case, the greatest degree of uncertainty in these analyses is introduced by the $\Lambda$ values, which are generally considered to be accurate to roughly 20 \%.  The uncertainty reported in Table 1 resulted from consideration of possible errors in all parameters, especially $\Lambda$.

\begin{figure}[b]

\includegraphics*[width=3.0 in]{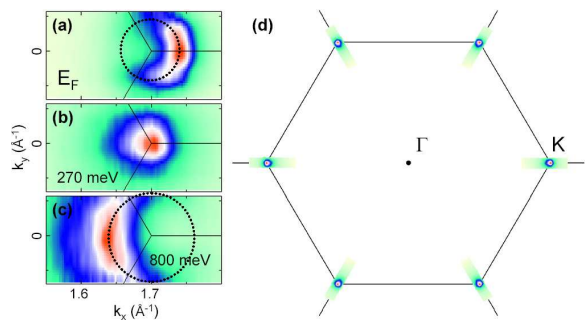}

\caption {High-resolution ARPES data taken on sample C\@. (a-c) Constant energy maps taken at $E_F$ (a) and finite binding energies (b,c) as indicated.  Dotted circles in (a) and (c) are guides to the eye. (d) Constant energy map of (b), symmetrized by 6-fold rotations.}

\label {fig4}

\end{figure}

Figure 4 shows high-resolution ARPES data in color maps, where intensity increases from light green to blue (half point between minimum and maximum) to white to orange. Panels (a) through (c) show constant energy intensity maps with $\approx 30$ meV energy window of integration for each binding energy. Also shown are Brillouin zones for graphite.  Note that the accuracy of our experiment does not allow for the determination a possible
minute difference \cite {charrier, forbeaux} in the lattice constant of our thin graphene layers relative to bulk graphite. 
The data in panels (a) through (c) are consistent with a band dispersion whose density of states is finite at $E_F$ [panel (a)], and goes through a decrease [panel (b)] and an increase [panel (c)] as binding energy increases. This behavior is roughly consistent with the behavior of a ``Dirac-cone'' band dispersion expected of a single graphene layer, albeit shifted in energy.  However, as we will discuss in a subsequent publication, there may be important differences as well. Panel (d) shows a 6-fold symmetrized map of the data in panel (b), demonstrating a 6-point-like constant energy map. To our knowledge, this is the best resolved such map reported for graphite so far.

Even so, we find that these maps are not limited by the instrumental momentum resolution, which is several times better than the width of the sharpest momentum structure [panel (a)]. Reading off the locations of the blue, the full width half maximum (FWHM) in momentum perpendicular to constant energy contour ($\sim$ circle) can be estimated as $\approx$ 0.05 \AA$^{-1}$.  If this width is due to a finite terrace size, then the latter is given by
$2 \pi /$FWHM $\approx$ 130 \AA, 
in rough agreement with the small length scale observed in Figure 2. Note also that this FWHM value is very similar to the width along the same direction reported in x-ray diffraction measurement \cite {charrier}, with a slightly better sharpness (20 \%). On the other hand, this FWHM value is much worse than the FWHM value observed in HOPG \cite {zhou} by a factor of 2, despite the fact that the map is now fully momentum resolved in contrast to the map only radially resolved for HOPG\@. While measurements on other thin graphene samples would be necessary to make a definite conclusion, we tentatively interpret this as an indication that the thin graphene samples carefully characterized here and in the literature \cite {charrier} are {\em not} better than HOPG in terrace (or grain) size.

In summary, we have succeeded in producing atomically-thin graphite films of thicknesses down to 1-2 graphene layers, grown epitaxially on the (0001) surface of 6H-SiC. Long recognized as a potentially ideal two-dimensional system for studies of novel electronic properties, graphene is also a sought-after material in relation to technological applications.  While the synthesis of truly perfect single-layer graphene sheets has not yet been realized, the availability of the thin graphene films characterized here nonetheless presents many exciting opportunities for future work in both carbon-based nanoelectronics and in the study of low-dimensional graphitic systems.

\begin {acknowledgments}

We would like to thank A. Bostwick for experimental help in sample preparation. We would also like to thank Giulia Lanzara for help in performing the SEM measurements. This work was supported by the U.S. DOE under contract No.~DEAC03-76SF00098; by the U.S. NSF through grant No.~DMR03-49361 and grant No.~DMR04-39768 and by the Sloan Foundation.  The ALS is supported by the U.S. DOE under contract No.~DE-AC02-05CH11231.
\vspace {-0.3cm}

\end {acknowledgments}



\end{document}